\documentclass[]{biometrika}

\usepackage{amsmath}

\usepackage{times}
\usepackage{bm}
\usepackage{natbib}
\usepackage{graphicx}
\usepackage{bbm}
\usepackage{amssymb}
\usepackage{url}
\usepackage{color}
\usepackage{multirow}
\usepackage[plain,noend]{algorithm2e}

\makeatletter
\renewcommand{\algocf@captiontext}[2]{#1\algocf@typo. \AlCapFnt{}#2} 
\def\@algocf@capt@plain{top}
\renewcommand{\algocf@makecaption}[2]{%
  \addtolength{\hsize}{\algomargin}%
  \sbox\@tempboxa{\algocf@captiontext{#1}{#2}}%
  \ifdim\wd\@tempboxa >\hsize
    \hskip .5\algomargin%
    \parbox[t]{\hsize}{\algocf@captiontext{#1}{#2}}
  \else%
    \global\@minipagefalse%
    \hbox to\hsize{\box\@tempboxa}
  \fi%
  \addtolength{\hsize}{-\algomargin}%
}
\makeatother

\def\pr{\operatorname{pr}}
\def\KL{\operatorname{KL}}

\def\exp{\operatorname{exp}}

\begin{document}



\markboth{E. F. Lock \and D. B. Dunson}{Shared kernel Bayesian screening}

\title{Shared kernel Bayesian screening}

\author{Eric F. Lock}
\affil{Division of Biostatistics, University of Minnesota, Minneapolis,  Minnesota 55455, U.S.A\email{elock@umn.edu}}
\author{\and David B. Dunson}
\affil{Department of Statistical Science, Duke University, Durham,  North Carolina 27708, U.S.A \email{dunson@duke.edu}}

\maketitle

\begin{abstract}

This article concerns testing for equality of distribution between groups.  We focus on screening variables with  shared distributional features such as common support, modes and patterns of skewness.  We propose a Bayesian testing method using kernel mixtures, which improves performance by borrowing information across the different variables and groups through shared kernels and a common probability of group differences.  The inclusion of shared kernels in a finite mixture, with Dirichlet priors on the weights, leads to a simple framework for testing that scales well for high-dimensional data.  We provide closed asymptotic forms for the posterior probability of equivalence in two groups and prove consistency under model misspecification.  The method is applied to DNA methylation array data from a breast cancer study, and compares favorably to competitors when type I error is estimated via permutation.  
\end{abstract}

\begin{keywords}
Epigenetics; Independent screening; Methylation array; Misspecification; Multiple comparisons; Multiple testing; Nonparametric Bayes inference. 
\end{keywords}

\section{Introduction}

\subsection{Motivation}
\begin{figure}[!ht]
	\centerline{\includegraphics[scale=0.8, trim = 0mm 0mm 0mm 0mm, clip = TRUE]{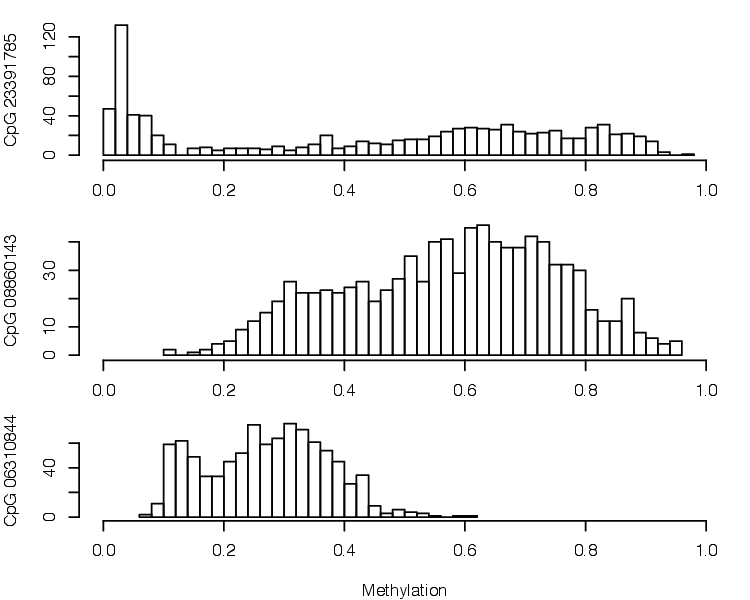}} 
	\caption{Distribution of methylation measurements at three CpG sites.  For each histogram the vertical scale gives frequency of occurrence among individuals.} \label{DistExamples}
\end{figure}

In modern biomedical research, it is common to screen for differences between groups in many variables.  These variables are often measured using the same technology and are not well characterized using a simple parametric distribution.  As an example we consider DNA methylation arrays.  Methylation is an epigenetic phenomenon that can affect transcription and occurs at genomic locations where a cytosine nucleotide is followed by a guanine nucleotide, called a CpG site.  High-throughput microarrays are commonly used to measure methylation levels for thousands of CpG sites genome-wide.  Measurements are typically collected from a tissue that contains several distinct cell types, and at a given CpG site each cell type is typically either methylated or unmethylated \citep{Reinius2012}.  Arrays therefore give continuous measurements for discrete methylation states and the resulting values are between $0$, no methylation, and $1$, fully methylated.  Figure~\ref{DistExamples} shows the distribution of methylation measurements over individuals for three CpG sites using data from the \citet{Perou2012}. Multi-modality and skewness are common; kernel mixtures are useful for modeling such complexity.
    
Methylation variables share several distributional features such as common support, common modes and common patterns of skewness.  The use of kernels that are shared across variables thus not only reduces computational burden but can also improve performance. It is also natural to share kernels across groups, with the interpretation that two groups arise from the same discrete process but in potentially different proportions.

We introduce a simple, computationally efficient, and theoretically supported Bayesian approach for screening using shared kernels across groups and, if appropriate, across variables.  The population distribution for each variable is approximated using a mixture of kernels $\{F_k\}_{k=1}^K$. For two groups $0$ and $1$, we test whether the groups have different kernel weights.  Specifically, for group distributions $F_m^{(0)}$ and $F_m^{(1)}$ at variable $m$, $F_m^{(0)} = \sum_{k=1}^K \pi_{mk}^{(0)} F_k$ and $F_m^{(1)} = \sum_{k=1}^K \pi_{mk}^{(1)} F_k$,  the competing hypotheses are
 \begin{align}H_{0m}: \pi_{mk}^{(0)} = \pi_{mk}^{(1)} \, \text{for all } k \, \, \, \text{vs} \, \,\,   H_{1m}: \pi_{mk}^{(0)} \neq \pi_{mk}^{(1)} \, \text{for some } k. \label{HypFramework}\end{align}
In practice $F_1,...,F_K$ and a shared Dirichlet prior distribution for the weights $\Pi_m^{(0)}, \Pi_m^{(1)}$ are estimated empirically.  A simple and tractable Gibbs sampling procedure is then used to estimate the posterior probability of $H_{0m}$ for each variable.  

While methylation array data provide excellent motivation, our framework addresses the general statistical problem of testing for equality between two groups that are drawn from the same strata but in potentially different proportions.  We argue that the method may also be useful for applications that do not have such a clear  interpretation, and this is supported with theoretical results in Section~\ref{consist}.

\subsection{Related work} 

The multi-modality of methylation measurements is widely recognized \citep{Laird2010} but often not accounted for in practice.  The two-sample t-test is most commonly used to identify sites of differential expression in case-control studies \citep{Bock2012}.  Alternative testing approaches are rank-based or discretize the data based on arbitrary thresholds \citep{Chen2011, Qiu2012}.  Other statistical models have been proposed to identify CpG sites that are hypomethylated, hypermethylated or   undifferentiated with respect to normal cells \citep{Khalili2007,Akalin2012}.  The focus on differential methylation levels between groups may miss other important differences between group distributions; for example, certain genomic regions have been shown to exhibit more variability in methylation, and thus epigenetic instability, among cancer cells than among normal cells \citep{Hansen2011}. 

Although our model involves finite mixtures, it is intended to be robust to parametric assumptions and so is comparable to nonparametric methods.  There is a literature on nonparametric Bayes testing of equivalence in distribution between groups.   \citet{DunsonPeddada2008} use a dependent Dirichlet process to test for equality against stochastically ordered alternatives.  They use an interval test based on total variation distance, and the framework is easily extended to unordered alternatives.  
\citet{Pennell2008} also use a Dirichlet process model for multiple groups and an interval test.  \citet{Ma2011} and \citet{Holmes2015} use Polya tree priors to test for exact equality.  Existing nonparametric Bayes tests do not exploit shared features among variables, in the form of shared kernels or otherwise. 

If kernel memberships are known, our testing framework (\ref{HypFramework}) is equivalent to a test for association with a $2 \times K$ contingency table. For this there are standard frequentist methods such as Fisher's exact test and Pearson's chi-square test, and established Bayesian methods \citep{Good1987,Albert1997}.   In our context the component memberships are unknown and are inferred probabilistically.  \cite{Xu2010} addressed this as part of a series of comparisons for Bayesian mixture distributions between groups.  They compare marginal likelihoods for models with and without assuming constant weights between groups.  Our focus is instead on screening settings in which there are many variables, and it is important to borrow information while adjusting for multiple testing.  Shared kernels facilitate borrowing of information and computational scaling, and in our implementation a shared prior for the probability of equality at each variable induces a multiplicity adjustment with favorable properties \citep{Scott2006,Muller2007,Scott2010}.

\section{Model}
\label{model}

\subsection{Shared kernel mixtures}
\label{SharedKern}
Below we describe the general model for shared kernel Bayesian screening.  Details that are specific to our implementation for methylation array data, including estimation techniques that facilitate posterior computation in high-dimensions, are given in Section~\ref{meth}.   

First we describe a shared kernel mixture model, to lay the groundwork for the two-group screening model in Section~\ref{TwoClass}.  Given data $x_{mn}$ for $M$ variables ($m=1,\hdots,M$) and $N$ subjects ($n=1,\hdots,N$), the shared kernel model assumes that observations $x_{mn}$ are realized from one of $K$ component distributions $F_1,\hdots,F_K$. Typically $x_{mn}$ is a continuous and unidimensional observation, but we present the model in sufficient generality to allow for more complex data structures.  We assume that$F_1,\hdots,F_K$ have corresponding likelihoods from the same parametric family $f(\cdot,\theta_k)$.  

Let $c_{mn} \in \{1,\hdots,K\}$ represent the component generating $x_{mn}$, and $\pi_{mk} = \pr (c_{mn} = k)$ be the probability that an arbitrary subject belongs to component $k$ in variable $m$. The generative model is $x_{mn} \sim F_k \text{ with probability } \pi_{mk}$. Under a Bayesian framework one puts a prior distribution on $\{\Pi_m = (\pi_{m1},\hdots,\pi_{mK})\}_{m=1}^M$ and, if they are unspecified, the kernels $F_1,\hdots,F_K$.  It is natural to use a Dirichlet conjugate prior for $\Pi_m$, characterized by a $K$-dimensional parameter $\alpha$ of positive real numbers.  Small values of $\alpha$, with $\alpha_k  \leq 1$, will favor small values for a subset of the $\pi_{mk}$ values.  Thus, some kernels may have negligible impact for a given variable.

\subsection{Two-group screening}
\label{TwoClass}
We extend the shared kernel model above to allow for two sample groups: $X^{(0)}$  with data $x_{mn}^{(0)}$ for $N_0$ subjects ($n=1,\hdots,N_0$; $m=1,\hdots,M$),  and 
$X^{(1)}$ with data $x_{mn}^{(1)}$ for $N_1$ subjects ($n=1,\hdots,N_1$; $m=1,\hdots,M$).  Observations for all $M$ variables are realized from a common set of kernels $F_1,\hdots,F_K$, but the two groups have potentially different weights $\{\Pi_m^{(0)}\}_{m=1}^M$ and $\{\Pi_m^{(1)}\}_{m=1}^M$.   

The weights $\Pi_m^{(0)}$ and $\Pi_m^{(1)}$ each have prior distribution Dir($\alpha$), whether they are identical or not.   Let $H_{0m}$ be the event that the mixing weights are the same for both groups:  $ \Pi_m^{(0)}= \Pi_m^{(1)}$.  Under $H_{1m}$, $\Pi_m^{(0)}$ and $\Pi_m^{(1)}$ are considered independent realizations from Dir($\alpha$).  Let $F_m^{(0)}$ be the distribution for group $0$ and let $F_m^{(1)}$ be the distribution for group $1$.  We consider a dummy variable $\mathbbm{1}(H_{0m}) \sim \text{Bernoulli}\{ \pr (H_{0m})\}$ and independent realizations $\tilde{\Pi}_{mk}, \tilde{\Pi}_{mk}^{(0)},\tilde{\Pi}_{mk}^{(1)} \sim \text{Dir}(\alpha)$ to give the joint distribution for groups $i=0,1$: 
\[F_m^{(i)} = \sum_{k=1}^K \big[ \mathbbm{1}(H_{0m}) \tilde{\pi}_{mk} + \{ 1-\mathbbm{1}(H_{0m}) \} \tilde{\pi}_{mk}^{(i)}\big] F_k .\]
As $\pr (H_{0m}) \rightarrow 1$,  $F_m^{(0)}$ and $F_m^{(1)}$ share the same mixing weights, and as $\pr (H_{0m}) \rightarrow 0$ the weights are independent.  

Let $\vec{n}_m^{(0)} = (n_{m1}^{(0)},\hdots,n_{mK}^{(0)})$ give the number of subjects in group $0$ that belong to each kernel $k$ in variable $m$, and define $\vec{n}_m^{(1)}$ similarly for group $1$.  Then, $\vec{n}_m= \vec{n}_m^{(0)}+\vec{n}_m^{(1)}$ gives the total number of subjects allocated to each component.  Under $H_{0m}$, the distribution for the component memberships $C_m^{(0)}$ and $C_m^{(1)}$ is 
\begin{align*}
\pr (C_m^{(0)},C_m^{(1)}\mid H_{0m}) & = \int_{\Pi} \pr (C_m^{(0)},C_m^{(1)}\mid\Pi) f(\Pi\mid\alpha) d \Pi \\
&= \frac{\Gamma(\sum_{k=1}^K \alpha_k)}{\Gamma(\sum_{k=1}^K n_{mk}+\alpha_k)} \prod_{k=1}^K \frac{\Gamma(n_{mk}+\alpha_k)}{\Gamma(\alpha_k)} \\
&= \beta(\vec{n}_m+\alpha) / \beta(\alpha),\label{deriv1}
\end{align*}
where $\Gamma$ is the gamma function and $\beta$ is the multivariate beta function $\beta(\alpha) = \prod_{k=1}^{K} \Gamma(\alpha_k)/\Gamma(\sum_{k=1}^K \alpha_k )$.
Similarly, under $H_{1m}$,
\begin{align*}
\pr (C_m^{(0)},C_m^{(1)}\mid H_{1m}) &= \int_{\Pi} \pr (C_m^{(0)}\mid\Pi) f(\Pi_m\mid\alpha) d \Pi   \int_{\Pi} \pr (C_m^{(1)}\mid\Pi) f(\Pi \mid \alpha) d \Pi \\
&= \frac{\beta(\vec{n}_m^{(0)}+\alpha)\beta(\vec{n}_m^{(1)}+\alpha)}{\beta(\alpha)^2}.
\end{align*}
Let the shared prior probability of no difference be $P_0 = \pr (H_{0m})$ for all $m$. The posterior probability of $H_{0m}$ given $C_m^{(0)}$ and $C_m^{(1)}$ is
\begin{equation}
\label{PostKnown}
  \begin{aligned}
\pr (H_{0m}\mid C_m^{(0)},C_m^{(1)}) &= \frac{P_0\, \pr (C_m^{(0)},C_m^{(1)}\mid H_{0m})}{P_0\, \pr (C_m^{(0)},C_m^{(1)}\mid H_{0m})+(1-P_0)\, \pr (C_m^{(0)},C_m^{(1)}\mid H_{1m})} \\
&=  \frac{P_0 \beta(\alpha)\beta(\vec{n}_m+\alpha)}{P_0 \beta(\alpha) \beta(\vec{n}_m+\alpha)+(1-P_0) \beta(\vec{n}_m^{(0)}+\alpha) \beta (\vec{n}_m^{(1)}+\alpha)},
\end{aligned}
\end{equation}

But in practice the kernel memberships are unknown, and the kernels may be unknown as well.  There is no analogous closed form that accounts for uncertainty in $(C_m^{(0)},C_m^{(1)})$  and direct computation is usually infeasible.  We instead employ a Gibbs sampling procedure that uses (\ref{PostKnown}) to approximate the full posterior distribution.  Under multiple related tests, $M>1$, we infer $P_0$ using a Be$(a,b)$ prior, where by default $a=b=1$.  The mean of the realized values of $\pr (H_{0m} \mid C_m^{(0)},C_m^{(1)})$ over the sampling iterations is used to estimate the posterior probability of $H_{0m}$ for each variable.   

 While this article focuses on the two-group case, extensions to multiple groups are straightforward.  A natural approach is to define a prior to cluster the groups.  For example, we could use a Dirichlet process as in \citet{Gopalan1998}, but instead of clustering group means we would be clustering group distributions.  Each cluster would then have a separate weight vector drawn from a Dirichlet distribution.

The above approach is presented in the context of shared kernels for high-dimensional screening, large $M$. The framework is also useful in the simple case $M=1$, and is particularly well motivated when two groups have the same strata but in potentially different proportions.  The theoretical results presented in Sections~\ref{asymp} and ~\ref{consist} are not specific to high-dimensional screening, and we drop the variable subscript $m$ for simplicity. 

\section{Asymptotic forms}
\label{asymp}
We investigate the asymptotic forms that result from Equation~(\ref{PostKnown}) as the number of observations tends to infinity.  Proofs are given in the Supplementary Material. 

Let $N = N_0+N_1$ and fix $\lambda_0 = N_0/(N_0 + N_1)$.  In Theorem~\ref{AsympRatio} we derive the asymptotic form of the conditional Bayes factor $\pr (H_0\mid C^{(0)},C^{(1)}) / \pr (H_1\mid C^{(0)},C^{(1)})$.

\begin{theorem}
\label{AsympRatio}
Let $\vec{p}_0 = \vec{n}^{(0)}/N_0$, $\vec{p}_1 = \vec{n}^{(1)}/N_1$, $\vec{p} = (\vec{n}^{(0)}+\vec{n}^{(1)})/N$, $r_{0k} = p_{0k}/p_k$ and $r_{1k} = p_{1k}/p_k$.  Then, as $N_0 , N_1 \rightarrow \infty$, 
\[\frac{\pr (H_0\mid C^{(0)},C^{(1)})}{\pr (H_1\mid C^{(0)},C^{(1)})} \sim c  N^{\frac{K-1}{2}}\prod_{k=1}^K r_{0k}^{-n_k^{(0)}}  r_{1k}^{-n_k^{(1)}}\]
where 
\[ c = \frac{P_0}{1-P_0} \left\{ \frac{\lambda_0 (1-\lambda_0)}{2\pi} \right\}^{\frac{K-1}{2}} \prod_{k=1}^K p_k^{\alpha_k + 1/2} (r_{0k}r_{1k})^{1/2-\alpha_k}.\]
\end{theorem}

The asymptotic form given in Theorem~\ref{AsympRatio} does not depend on the generative distribution.  In the following we consider corollaries under $H_0$ and $H_1$.  
\begin{corollary}
\label{AsympH0}
Under  $H_0: \Pi^{(0)}=\Pi^{(1)}=\Pi$,
\[ \frac{\pr (H_0\mid C^{(0)},C^{(1)})}{\pr (H_1\mid C^{(0)},C^{(1)})} \sim c  N^{\frac{K-1}{2}}  \prod_{k=1}^K  \exp\left\{-\frac{\{\lambda_0(1-\lambda_0)\}^{1/2}}{2\pi_k} N(p_{0k}-p_{1k})^2 \right\},\]
where  \[\{\lambda_0(1-\lambda_0)\}^{1/2} N(p_{0k}-p_{1k})^2 \sim \chi_1^2.\]
\end{corollary}

It follows that under $H_0$ the log Bayes factor has order $\frac{1}{2}(K-1)\text{log}(N) + O_p(1)$, and therefore $\pr (H_0\mid C^{(0)},C^{(1)})$ converges to $1$ at a sublinear rate.  

\begin{corollary}
\label{AsympH1}
Under $H_1: \Pi^{(0)} \neq \Pi^{(1)}$, let $\Pi^* = \lambda_0 \Pi^{(0)}+(1-\lambda_0)\Pi^{(1)}$. Then,
\[\frac{\pr (H_0\mid C^{(0)},C^{(1)})}{\pr (H_1\mid C^{(0)},C^{(1)})} \sim c N^{\frac{K-1}{2}} \prod_{k=1}^K \left(\frac{\pi_k^{(0)}}{\pi_k^*}\right)^{-N\lambda_0\pi_k^{(0)}} \left(\frac{\pi_k^{(1)}}{\pi_k^*}\right)^{-N (1-\lambda_0) \pi_k^{(1)}} \exp \left\{O_p \left(N^{1/2}\right) \right\}.\]
\end{corollary}

It follows that under $H_1$ the log Bayes factor has order \[- N \sum \bigg\{ \lambda_0 \pi_{k}^{(0)} \log\bigg(\frac{\pi_k^{(0)}}{\pi_k^*}\bigg)+(1-\lambda_0) \pi_{k}^{(1)} \log\bigg(\frac{\pi_k^{(1)}}{\pi_k^*}\bigg) \bigg\}+ O_p \left(N^{1/2}\right),\] and therefore $\pr (H_0\mid C^{(0)},C^{(1)})$ converges to zero at an exponential rate.

Exponential convergence under $H_1$ and sublinear convergence under $H_0$ has been observed for many Bayesian testing models \citep{Kass1995, Walker2004}.  \citet{Johnson2010} discuss this property for local prior densities, in which regions of the parameter space consistent with $H_0$ also have non-negligible density under $H_1$;  they give general conditions for which the Bayes factor has order $N/2$ under $H_0$ and converges exponentially under $H_1$ when testing a point null hypothesis for a scalar parameter.  In our view the asymmetry in asymptotic rates under $H_0$ and $H_1$ is reasonable in our case and in most other models, as $H_0$ is much more precise.  In practice, we still obtain strong evidence in favour of $H_0$ for moderate samples.

The exact asymptotic distributions given in Corollaries~\ref{AsympH0} and~\ref{AsympH1} are derived under the assumption that the component memberships  $C^{(0)}$ and $C^{(1)}$ are known, but in practice they are unknown.  Additionally, the component distributions $F_1,\hdots,F_K$ may be unknown.  A simulation study presented in the Supplemental Material suggests that the asymptotic rates derived above also hold with a prior on $C^{(0)}$, $C^{(1)},$ and $F_1,\hdots,F_K$.  
\section{Consistency}
\label{consist}
We establish consistency of the method as a test for equality of distribution under very general conditions.  The following results allow for misspecification in that the true data generating distribution may not fall within the support of the prior.  For example, $F^{(0)}$ and $F^{(1)}$ may not be finite mixtures of simpler component distributions.  Such misspecified models clearly do not provide a consistent estimator for the full data generating distribution, but, as we show, they can still be consistent as a test for equality of distribution.  Proofs for all theorems, corollaries, and remarks in this section are given in Appendix~\ref{Proofs}.

First, we derive asymptotic results for a one-group finite mixture model under misspecification.  The proof of our first result uses general asymptotic theory for Bayesian posteriors under misspecification given in \cite{Kleijn2006}, and we borrow their notation where appropriate.  Theorem~\ref{SingMixtureCont} below implies that the posterior for a mixture distribution will converge to the convex combination of component distributions $f^*$ that is closest in terms of Kullback--Leibler  divergence to the true density $f_0$.  First, we define $B(\epsilon, f^*; f_0)$ to be a neighborhood of the density $f^*$ under the measure induced by the density $f_0$: 
\[B(\epsilon, f^*; f_0) = \left \{ f \in \mathbb{F}: -\int f_0  \log \frac{f}{f^*} \leq \epsilon^2, \int f_0 \left( \log \frac{f}{f^*} \right)^2 \leq \epsilon^2 \right \},\]
and define $d(f_1,f_2)$ to be the weighted Hellinger distance  \[ d^2 (f_1,f_2) = \frac{1}{2} \int (f_1^{1/2}-f_2^{1/2})^2 \frac{f_0}{f^*}.\]

\begin{theorem}
\label{SingMixtureCont}
Let $x_1, \hdots, x_N$ be independent with density $f_0$.  Let $\mathbb{F}$ be the set of all convex combinations of dictionary densities $\{f_k\}_{k=1}^{K}$, and let $P$ define a prior on $\mathbb{F}$.  Assume  $f^*= \underset{f \in \mathbb{F}}{\text{argmin}} \, \KL (f_0 || f^*)$ exists and $\pr\{ B(\epsilon,f^*;f_0)\}>0$ for all $\epsilon >0$.  Then, for any fixed $\epsilon>0$,
\[\pr \{ f \in \mathbb{F}: d(f,f^*) \geq \epsilon \mid x_1, \hdots, x_N\} \rightarrow 0.\]  
\end{theorem}

The prior support condition $\pr\{ B(\epsilon,f^*;f_0)\} >0$ for all $\epsilon>0$ is satisfied for all priors that have positive support over $\mathbb{F}$.  This includes priors for $\Pi$ with positive support over the unit simplex $\mathbb{S}^{K-1}$, such as Dirichlet priors.  Although the weighted Hellinger distance $d$ is non-standard, convergence in $d$ implies convergence of the component weights, as shown in Corollary~\ref{PiCons}. 

\begin{corollary}
\label{PiCons}
 Under the setting of Theorem~\ref{SingMixtureCont}, let $\Pi^*=(\pi_1^*,\hdots,\pi_K^*)$ be the component weights corresponding to $f^*$.  Assume $\Pi^*$ is unique in that $\sum \pi_k f_k = \sum \pi_k^* f_k = f^*$
only if $\Pi=\Pi^*$.  Then, for any fixed $\epsilon>0$,
\[\pr (\Pi \in \mathbb{S}^{K-1}: ||\Pi-\Pi^*|| \geq \epsilon \mid x_1, \hdots, x_N) \rightarrow 0.\]
\end{corollary}

Uniqueness of the component weights at $f^*$ is trivially satisfied if distinct mixture weights yield distinct distributions in $\mathbb{F}$. Such identifiability has been established in general for Gaussian mixtures with variable means and variances, as well as for several other common cases \citep{Teicher1963,Yakowitz1968}. 

Kullback--Leibler  divergence over $\mathbb{F}$ is convex, and its minimizer $f^*$ satisfies interesting conditions.

\begin{remark}
\label{fstarcond}
Under the setting of Theorem~\ref{SingMixtureCont}, assume $\pi_k^* > 0 $ for $k=1,\hdots,K$ and $\sum \pi_k^*=1$.  Then, $f^* = \sum \pi^*_k f_k$ achieves the minimum Kullback--Leibler divergence in $\mathbb{F}$ with respect to $f_0$ if and only if
\[\int \frac{f_1}{f^*} f_0 = \cdots =  \int \frac{f_K}{f^*} f_0.\]
If some $\pi_k^*=0$, the minimum divergence is achieved where $\int (f_k / f^*) f_0$ are equivalent for all $\pi_k^*>0$.   
\end{remark}

We now give the result for consistency as a test for equality of distribution.  

\begin{theorem}
\label{ConsTheorem}
Assume $x_1^{(0)},\hdots, x_{N_0}^{(0)}$ are independent with density $f^{(0)}$, $x_1^{(1)},\hdots, x_{N_1}^{(1)}$ are independent with density $f^{(1)}$, and let
\[f^{* (0)} = \underset{f \in \mathbb{F}}{\text{argmin}} \, \KL (f^{(0)}|| f) \, \, , \,\, f^{* (1)} = \underset{f \in \mathbb{F}}{\text{argmin}} \, \KL (f^{(1)}|| f).\]
Assume the uniqueness condition for Corollary~\ref{PiCons} holds for $f^{*(0)}$ and $f^{*(1)}$.  If $f^{ (0)} = f^{ (1)}$,  $\pr (H_0 \mid X) \rightarrow 1$ as $N \rightarrow \infty$.
  If $f^{* (0)} \neq f^{* (1)}$,  $\pr (H_0 \mid X) \rightarrow 0$ as $N \rightarrow \infty$.
\end{theorem}

Theorem~\ref{ConsTheorem} implies that the posterior probability of equality is consistent under $H_0$, even under misspecification.  Consistency under $H_1$ holds generally under misspecification, but fails if $f^{(0)}$ and $f^{(1)}$ are both closest in Kullback--Leibler  divergence to the same $f^* \in \mathbb{F}$.  This can occur if $f^{(0)}$ and $f^{(1)}$ are both closer to the same component distribution $f_k$ than they are to any other distribution in the convex hull.  

\section{Application to methylation data}
\label{meth}

\subsection{Data and Estimation}
\label{dict}
We illustrate our approach on a methylation array dataset for $N=597$ breast cancer samples and $M=21,986$ CpG sites.  These data are publicly available from the TCGA Data Portal \citep{Perou2012}.   We focus on testing for a difference between tumours that are identified as basal-like ($N_0 = 112$) against those that are not ($N_1 = 485$) at each site.   Basal-like samples have a relatively poor clinical prognosis and a distinct gene expression profile, but the role of DNA methylation in this distinction has not been well characterized.  

For scalability and to borrow information across sites, we apply a two-stage procedure.    First, a set of dictionary kernels are estimated.  Specifically, for $k=1,\hdots,K$, $f_k$ is the density of a normal distribution with mean $\mu_k$ and precision $\tau_k$ truncated to fall within the interval $[0,1]$. We use a normal-gamma prior for $\mu_k$ and $\tau_k$.  For computational reasons we estimate the posterior for $f_1,\hdots,f_K$ from a sub-sample of $500$ sites, for an effective sample size of $597 \times 500 = 298,500$ observations. We employ a Gibbs sampler and update the common Dirichlet prior parameter $\alpha$ at each iteration using maximum likelihood estimation \citep{Ronning1989}.  Alternatively one could use a hyperprior for $\alpha$, but this complicates posterior estimation and probably has little influence on posterior estimates as the effective sample size is very large. Similarly, we find there is little uncertainty in the posterior mean and variance for each kernel;  we can ignore the error in estimating these densities and fix them in the second stage.  

\begin{figure}[!ht]
	\centerline{\includegraphics[scale=0.5, trim = 0mm 0mm 0mm 0mm, clip = TRUE]{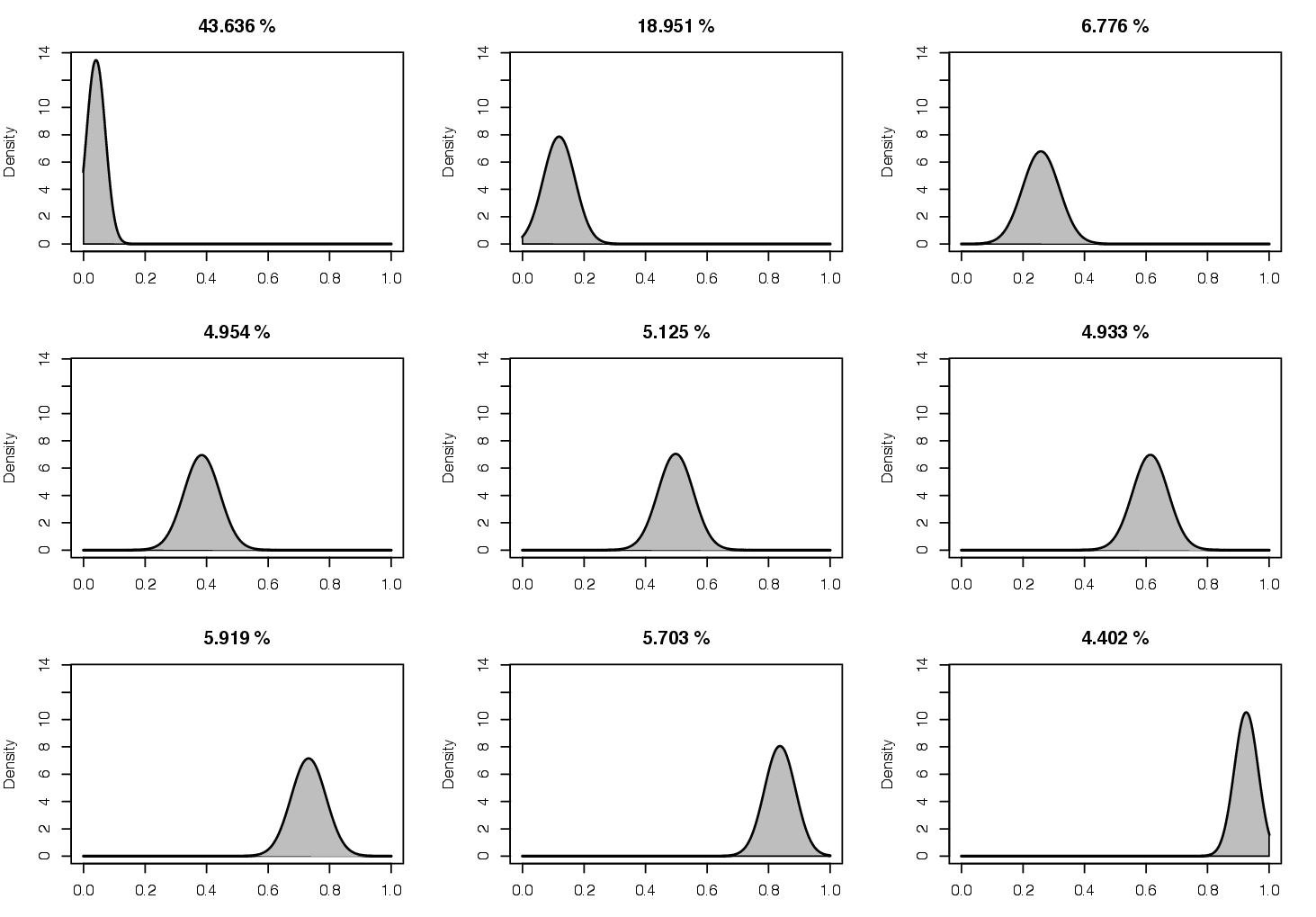}} 
	\caption{Truncated normal dictionary densities for $K=9$,  with the percentage of samples allocated to each density (over all sites).} \label{DictDensities}
\end{figure}

The number of kernels $K=9$ is chosen by cross validation based on the mean log-likelihood for held-out observations.  Estimates for the dictionary densities $f_1,\hdots,f_9$ are shown in Figure~\ref{DictDensities}; to address the label switching problem, we order the kernels by their means and then average over Gibbs samples.  For fixed $f_1,\hdots,f_9$, we compute the posterior for the two-group model at each CpG site using a simple and efficient Gibbs sampler and a uniform hyperprior for $P_0$.  We calculate the component likelihoods $f_k(x_{mn})$
for all sites $m$, samples $n$, and components $k$ in advance, which greatly reduces the computational burden.  

\subsection{Results}

We run Gibbs sampling for the two-group model for all $21,986$ CpG sites, with $5000$ iterations, after a $1000$ iteration burn-in.  The draws mix well and converge quickly; mixing is considerably improved by fixing the dictionary densities.

\begin{figure}[!ht]
	\centerline{\includegraphics[scale=0.6, trim = 0mm 0mm 0mm 0mm, clip = TRUE]{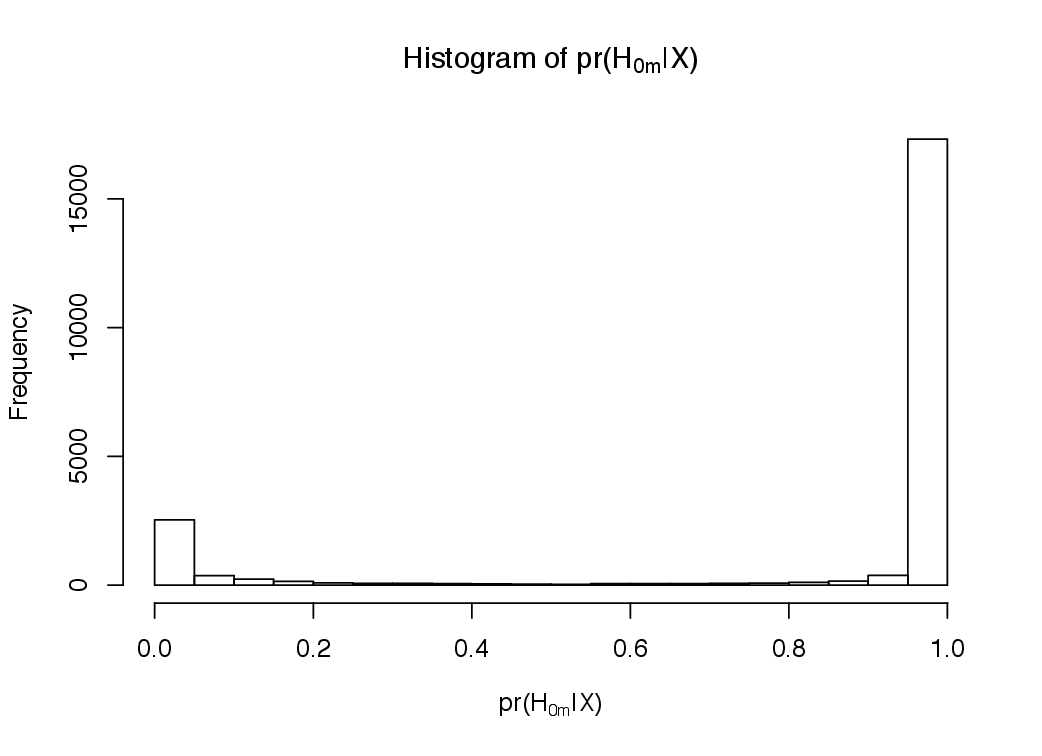}} 
	\caption{Histogram of posterior probabilities of $H_0$ at 21,986 CpG sites with $N_0=112$ basal and $N_1 = 485$ non-basal tumours.}  \label{PostProbHist}
\end{figure}

The global prior probability of no difference, inferred using a uniform hyperprior, was $\hat{P}_0 = 0.821$.
The estimated posterior probabilities $\pr (H_{0m} \mid X)$ are shown in Figure~\ref{PostProbHist}.  These have a U-shaped distribution, with 91\% of values falling below $0.05$ or above $0.95$.  Many values are close to 1, suggesting that these methylation sites play no role in the distinction between basal and non-basal tumours.    

\begin{figure}[!ht]
	\centerline{\includegraphics[scale=0.75, trim = 0mm 0mm 0mm 0mm, clip = TRUE]{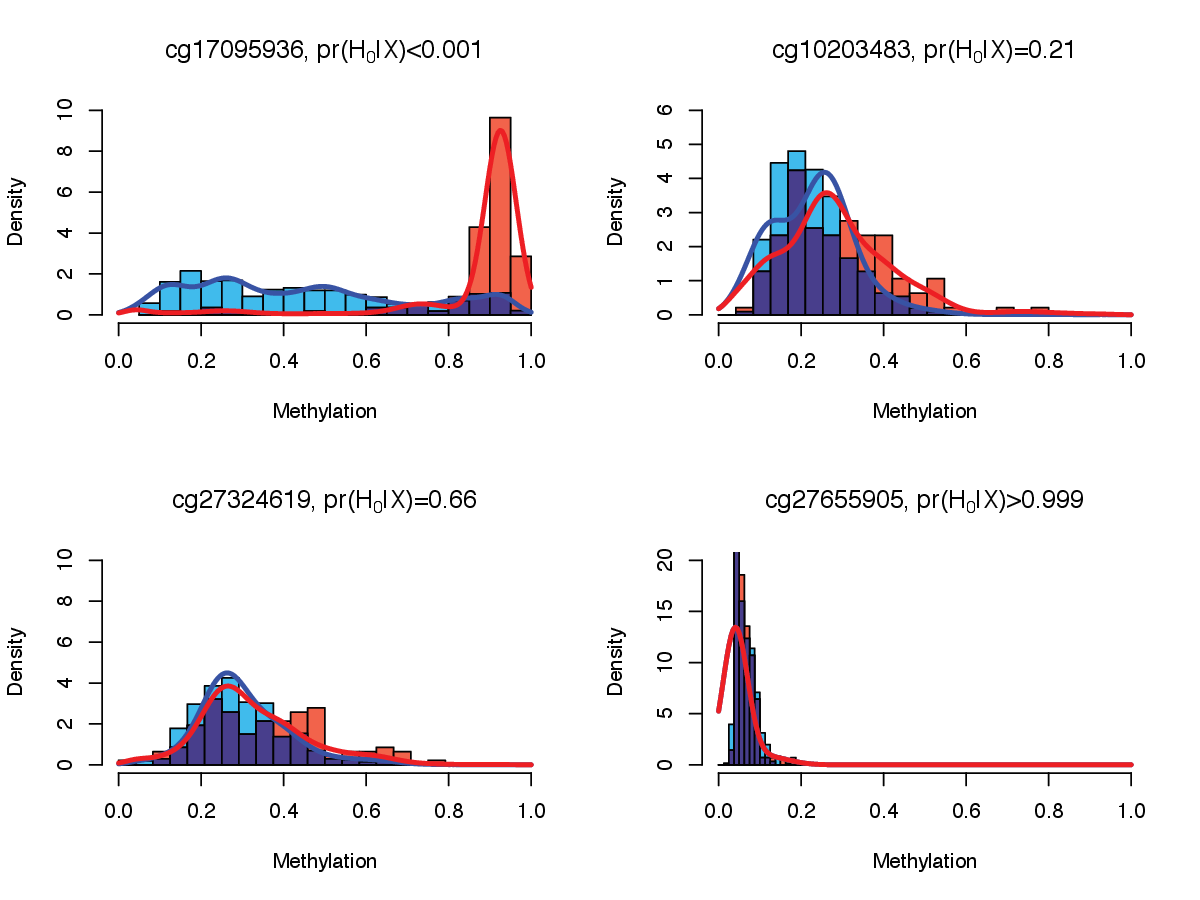}} 
\caption{ The estimated densities for basal (red) and non-basal (blue) samples for four CpG sites with different posterior probabilities of $H_0$.  Histograms are shown for both groups and their overlap is coloured violet.} \label{DictTestingFits}
\end{figure}

Figure~\ref{DictTestingFits} shows the sample distributions and mixture density fits for basal and non-basal tumours at four CpG sites.  These four sites were selected to show a range of estimated differences between the distributions for basal and non-basal tumours.  In general, the estimated mixture densities appear to fit the data well.  Some CpG sites with posterior probabilities $\pr (H_{0m} \mid  X)$ that are very small have dramatically different distributions between the two groups.  For the majority of CpG sites, the estimated distributions for the two groups are nearly identical.  The method naturally borrows strength across groups to estimate a common density when $\pr (H_{0m} \mid X)\rightarrow1$, and estimates the two densities separately when $\pr (H_{0m} \mid X)\rightarrow 0$.

We investigated the potential relevance of differentially distributed CpG sites by considering the expression of the gene at their genomic location. DNA methylation is thought to primarily inhibit transcription and therefore repress gene expression. Of 2117 CpG sites with $\pr(H_{0m} \mid X) < 0.01$, 1256 have a significant negative association with gene expression using Spearman’s rank correlation, p-value $< 0.01$.  For these cases methylation gives a potential mechanistic explanation for well-known differences in gene transcription levels between basal and non-basal tumours. In particular, these include five genes from the well-known PAM50 gene signature for breast cancer subtyping \citep{Parker2009}: \emph{MYBL2}, \emph{EGFR}, \emph{MIA}, \emph{SFRP1} and \emph{MLPH}.  
A supplemental spreadsheet gives the posterior probability $\pr(H_{0m} \mid X)$ and corresponding gene expression statistics for all CpG sites.

\section{Methods comparison on methylation data}
\label{MethodCompare}
We use data from Section~\ref{meth} to compare the power of testing methods on methylation array data.  We consider the following methods:
(a) the shared kernel test, as implemented in Section~\ref{meth} but with $P_0$ fixed at $0.5$ so that Bayes factors are independent,
(b) the two-sample Anderson--Darling test \citep{Scholz1987},
(c) a dependent optional Polya tree test \citep{Ma2011}, using code provided by the authors under default specifications,
(d) a Polya tree test (Holmes et al.), using code provided by the authors under default specifications, 
(e) the Wilcoxon rank sum test,
(f) the two-sample t-test with unequal variance,
(g) a restricted dependent Dirichlet process test \citep{DunsonPeddada2008}, using the interval null hypothesis $d_{\text{TV}} \in [0,0.05]$, where $d_{\text{TV}}$ is total variation distance.  Methods (a)--(d) are general tests for equality of distribution, while methods (e)--(g) test for different levels of methylation.       

\begin{figure}[!ht]
	\centerline{\includegraphics[scale=0.77, trim = 0mm 5mm 0mm 20mm, clip = TRUE]{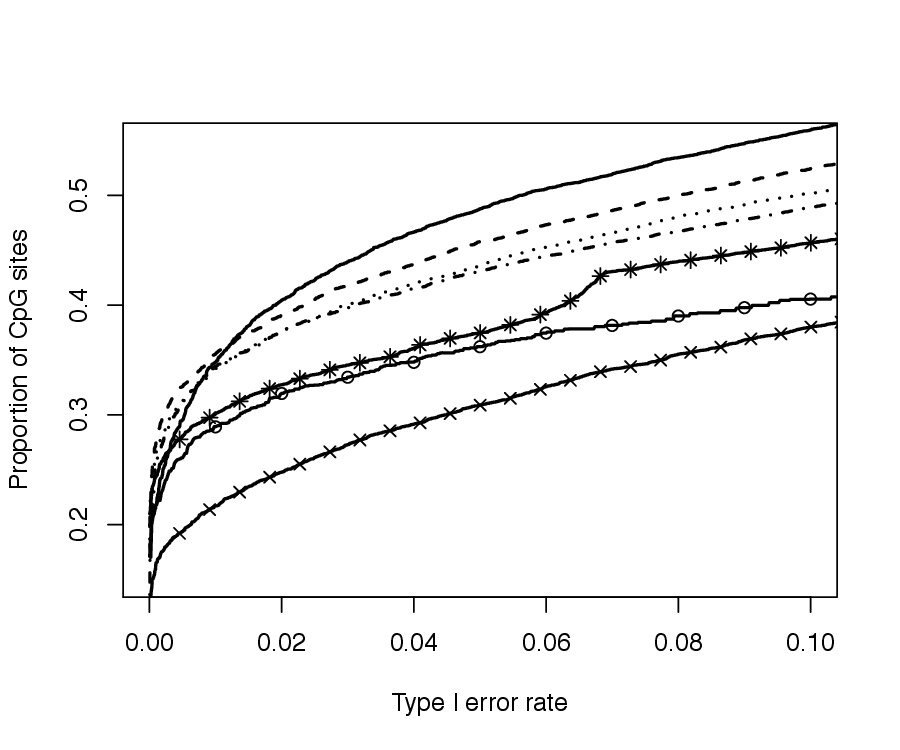}}
	\caption{Proportion of CpG sites identified as different between groups as a function of the proportion of sites identified as different under permutation, type I error rate, for seven different testing methods.  The curve for the shared kernel test is solid, for the Anderson-Darling test is dashed, for the t-test is dotted, for the Wilcoxon test is dot-dashed, for the dependent optional Polya tree test is starred, for the restricted dependent Dirichlet process test is circled, and for the Polya tree test is crossed.}  	\label{MethodComp}
\end{figure}

 We apply each method to test for a difference between basal and non-basal tumours at all $21,986$ CpG sites. For comparison, we also apply each method under random permutation of the group labels separately at each site to generate a null distribution.  The curves shown in Figure~\ref{MethodComp} are obtained by varying the threshold on the Bayes factor or p-value, depending on the method.  We compare the proportion of the $21,986$ CpG sites that are identified as different with the proportion of sites that are identified as different under permutation.  The proportion under permutation gives a robust estimate of the type I error rate, and so this is a frequentist approach to assessing discriminatory power.  The shared kernel test exceeds other Bayesian non-parametric tests by a wide margin.  It also generally performs as well or better than frequentist approaches, although the Anderson--Darling test is competitive.  Unlike nonparametric frequentist competitors the shared kernel approach admits a full probability model to assess strength of evidence for both the null and alternative hypotheses, which can be used in larger Bayesian models. Moreover, the shared kernel approach facilitates interpretation by modelling the full distribution, with uncertainty, for each group.     

\section*{Acknowledgment} 
This work was supported in part by grants from the National Institutes of Health.

\appendix

\appendixone
\section{Proofs}
\label{proofs}

\subsection{Proof of Theorem 1}
\label{AsympRatioProof}
\begin{proof}
Consider
\begin{align}
\frac{\beta(\vec{n})}{\beta(\vec{n}^{(0)})\beta(\vec{n}^{(1)})} &=\frac{\Gamma(N_0) \Gamma(N_1)}{\Gamma(N)} \prod_{k=1}^K \frac{\Gamma(n_k)}{\Gamma(n_k^{(0)})\Gamma(n_k^{(1)})} \nonumber \\
&= \frac{\beta(N_0,N_1)}{\prod_{k=1}^K\beta(n_k^{(0)},n_k^{(1)})}. \nonumber
\end{align}
Stirling's approximation gives 
\[ \beta(x,y) \sim (2\pi)^{1/2} x^{x-\frac{1}{2}}y^{y-\frac{1}{2}} (x+y)^{\frac{1}{2}-x-y},\]
and so 
\begin{align}
 \frac{\beta(\vec{n})}{\beta(\vec{n}^{(0)})\beta(\vec{n}^{(1)})} &\sim \frac{(2\pi)^{1/2} N_0^{N_0-\frac{1}{2}}N_1^{N_1-\frac{1}{2}} N^{\frac{1}{2}-N}}{\prod_{k=1}^K (2\pi)^{1/2} (p_{0k} N_0)^{p_{0k} N_0-\frac{1}{2}}(p_{1k} N_1)^{p_{1k}N_1-\frac{1}{2}} (p_k N)^{\frac{1}{2}-p_k N}} \nonumber \\
&=  \frac{(2\pi)^{1/2} N_0^{N_0-\frac{1}{2}}N_1^{N_1-\frac{1}{2}} (N)^{\frac{1}{2}-N}}{ (2\pi)^{\frac{K}{2}}  N_0^{N_0-\frac{K}{2}}N_1^{N_1-\frac{K}{2}} N^{\frac{K}{2}-N} \prod_{k=1}^K r_{0k}^{p_{0k}N_0-1/2} r_{1k}^{p_{1k}N_1-1/2} p_k^{-1/2}} \nonumber \\
&= \left( \frac{N_0 N_1}{2 \pi N} \right)^{\frac{K-1}{2}}\prod_{k=1}^K  r_{0k}^{1/2-n_k^{(0)}} r_{1k}^{1/2-n_k^{(1)}} p_k^{1/2}. \label{SterlResult}
\end{align}
Also 
\begin{align}
\beta(\vec{n}+\alpha) &= \frac{\prod_{k=1}^K \Gamma(p_k N+\alpha_i)}{\Gamma(N+\sum_{k=1}^K \alpha_k)} 
\sim \frac{\prod_{k=1}^K (p_k N)^{\alpha_k} \Gamma(p_k N)}{N^{\sum \alpha_k} \Gamma(N)} \label{usefact} \\ 
&= \frac{N^{\sum \alpha_k} \prod_{k=1}^{K} p_k^{\alpha_k}}{N^{\sum \alpha_k}} \beta(\vec{n}) 
=  \beta(\vec{n}) \prod_{k=1}^{K} p_k^{\alpha_k} \label{AsymBn} 
\end{align}
where  (\ref{usefact}) uses the approximation $\Gamma(x+c) \sim x^c \Gamma(x)$ as $x \rightarrow \infty$.  Similarly, 

\begin{align} \beta(\vec{n}^{(0)}+\alpha) \sim \beta(\vec{n}^{(0)}) \prod_{k=1}^{K} p_{0k}^{\alpha_k} \, \, \, \, \,
 \text{and}  \, \, \, \, \, \beta(\vec{n}^{(1)}+\alpha) \sim \beta(\vec{n}^{(1)}) \prod_{k=1}^{K} p_{1k}^{\alpha_k}. \label{AsymB01}\end{align}

Therefore
\begin{align}
 \frac{\beta(\vec{n}+\alpha)}{\beta(\vec{n}^{(0)}+\alpha)\beta(\vec{n}^{(1)}+\alpha)} &\sim \frac{\beta(\vec{n})}{\beta(\vec{n}^{(0)})\beta(\vec{n}^{(1)})}  \prod_{k=1}^{K} \left(\frac{p_{k}}{p_{0k}p_{1k}}\right)^{\alpha_k} \label{step1}\\
&\sim \left( \frac{N_0 N_1}{2 \pi N} \right)^{\frac{K-1}{2}}\prod_{k=1}^K p_k^{\alpha_k + 1/2} r_{0k}^{1/2-n_k^{(0)}-\alpha_k}  r_{1k}^{1/2-n_k^{(1)}-\alpha_k} \label{step2}
\end{align}
Where (\ref{step1}) follows from (\ref{AsymBn}) \& (\ref{AsymB01}), and (\ref{step2}) follows from (\ref{SterlResult}).  It follows from (\ref{step2}) and the full conditional given in Equation (2) of the main article that
\begin{align*}
\frac{\pr (H_0|C^{(0)},C^{(1)})}{\pr (H_1|C^{(0)},C^{(1)})} &= \frac{\beta(\alpha)P_0}{\pr (H_1)}   \frac{\beta(\vec{n}+\alpha)}{\beta(\vec{n}^{(0)}+\alpha)\beta(\vec{n}^{(1)}+\alpha)}  \\
&\sim  c  N^{\frac{K-1}{2}}\prod_{k=1}^K r_{0k}^{-n_k^{(0)}}  r_{1k}^{-n_k^{(1)}}
\end{align*}

where 
\[ c = \frac{P_0}{1-P_0} \left\{ \frac{\lambda_0 (1-\lambda_0)}{2\pi} \right\}^{\frac{K-1}{2}} \prod_{k=1}^K p_k^{\alpha_k + 1/2} (r_{0k}r_{1k})^{1/2-\alpha_k}.\]

\end{proof}

\subsection{Proof of Corollary 1}
\label{AsympH0proof}
\begin{proof}
Let $\lambda_1=1-\lambda_0$ and consider 
\begin{align*}
\log ( r_{0k}^{-n_k^{(0)}}  r_{1k}^{-n_k^{(1)}})  = N\{ p_k \log(p_k)- \lambda_0 p_{0k} \log(p_{0k})- \lambda_1 p_{1k}\log(p_{1k})\}.
\end{align*}
Under $H_0$, as $N \rightarrow \infty$, 
\[p_{0k} = \pi_k+\frac{Z_0}{(N\lambda_0)^{1/2}}\,\,\,\,\, \text{and} \,\,\,\,\, p_{1k} = \pi_k+\frac{Z_1}{(N\lambda_1)^{1/2}},\]
where $Z_0$, $Z_1$ are independent standard normal variables. 
It follows that $\log( r_{0k}^{-n_k^{(0)}}  r_{1k}^{-n_k^{(1)}}) =A+B$, where
\begin{align*}
A &= N \pi_k \left\{ \log \left(\pi_k+\frac{\lambda_0^{1/2}Z_0}{N^{1/2}}+\frac{\lambda_1^{1/2}Z_1}{N^{1/2}}\right)-\lambda_0 \log \left(\pi_k+\frac{Z_0}{(N\lambda_0)^{1/2}}\right)-\lambda_1 \log \left(\pi_k+\frac{Z_1}{(N\lambda_1)^{1/2}} \right)\right\} \\
&= N\pi_k \left\{ \log \left(1+\frac{\lambda_0^{1/2} Z_0}{\pi_k N^{1/2}}+\frac{\lambda_1^{1/2} Z_1}{\pi_k N^{1/2}} \right)-\lambda_0 \log \left(1+\frac{Z_0}{\pi_k (N\lambda_0)^{1/2}} \right)- \lambda_1 \log \left(1+\frac{Z_1}{\pi_k (N \lambda_1)^{1/2} } \right)\right\}
\end{align*}
and 
\begin{align*}
B =& \left\{(N\lambda_0)^{1/2}Z_0+(N\lambda_1)^{1/2}Z_1 \right\}\log \left(\pi_k+\frac{\lambda_0^{1/2} Z_0}{N^{1/2}}+\frac{\lambda_1^{1/2} Z_1}{ N^{1/2}} \right) \\&-(N\lambda_0)^{1/2} Z_0 \log \left(\pi_k+\frac{Z_0}{(N\lambda_0)^{1/2}} \right)- (N\lambda_1)^{1/2} Z_1 \log \left(\pi_k+\frac{Z_1}{(N \lambda_1)^{1/2}} \right) \\
=& \left \{ (N\lambda_0)^{1/2}Z_0+(N\lambda_1)^{1/2}Z_1 \right \}\log \left(1+\frac{\lambda_0^{1/2}Z_0}{\pi_k N^{1/2}}+\frac{\lambda_1^{1/2}Z_1}{\pi_k N^{1/2}} \right) \\&-(N\lambda_0)^{1/2}Z_0 \log \left(1+\frac{Z_0}{\pi_k (N\lambda_0)^{1/2}} \right)- (N\lambda_1)^{1/2}Z_1 \log \left(1+\frac{Z_1}{\pi_k (N \lambda_1)^{1/2}} \right).
\end{align*}
The Maclaurin expansion $\log(1+x)=\sum_{i=1}^\infty (-1)^{1+i} (x^i / i)$ gives
\begin{align*}
A &= 0 - \frac{1}{2 \pi_k} \left\{ \left( \lambda_0^{1/2} Z_0+\lambda_1^{1/2} Z_1 \right)^2-Z_0^2-Z_1^2 \right\} +O_p \left( N^{-\frac12} \right) \\
&= \frac{1}{2\pi_k} (\lambda_1^{1/2}Z_0-\lambda_0^{1/2}Z_1)^2 +O_p \left( N^{-\frac12} \right)
\end{align*}
and 
\begin{align*}
B &= \frac{1}{\pi_k} \left\{ (\lambda_0^{1/2}Z_0+\lambda_1^{1/2}Z_1)^2-Z_0^2-Z_1^2 \right\} +O_p \left( N^{-\frac12} \right) \\
&= -\frac{1}{\pi_k} (\lambda_1^{1/2}Z_0-\lambda_0^{1/2}Z_1)^2 +O_p \left( N^{-\frac12} \right).
\end{align*}
Thus, 
\begin{align*}
\log( r_{0k}^{-n_k^{(0)}}  r_{1k}^{-n_k^{(1)}}) &=A+B \\
&= - \frac{1}{2\pi_k} (\lambda_1^{1/2}Z_0-\lambda_0^{1/2}Z_1)^2 +O_p \left( N^{-\frac12} \right) \\
&= - \frac{(\lambda_0 \lambda_1)^{1/2}}{2\pi_k} N (p_0-p_1)^2 +O_p \left( N^{-\frac12} \right).
\end{align*}
 It follows from Theorem~ 1 that 
 \[ \frac{\pr (H_0|C^{(0)},C^{(1)})}{\pr (H_1|C^{(0)},C^{(1)})} \sim c   N^{\frac{K-1}{2}}  \prod_{k=1}^K  \exp \left\{-\frac{(\lambda_0 \lambda_1)^{1/2}}{2\pi_k} N(p_{0k}-p_{1k})^2 \right\}\]
as $N \rightarrow \infty$, where $(\lambda_0 \lambda_1)^{1/2} N (p_0-p_1)^2 \sim \chi_1^2$.
\end{proof}

\subsection{Proof of Corollary 2}
\label{AsympH1proof}
\begin{proof}
Let $\lambda_1=1-\lambda_0$,  and $\pi_k^* = \lambda_0 \pi_k^{(0)}+\lambda_1 \pi_k^{(1)}$.  As $N \rightarrow \infty$, 
\[p_{0k} = \pi_k^{(0)}+O_p \left( N^{-\frac12} \right), \,\,\,\,\, \,\,\,\,\, p_{1k} = \pi_k^{(1)}+O_p \left( N^{-\frac12} \right) \,\,\,\,\, \text{and} \,\,\,\,\, p_{k}^* = \pi_k^{*}+O_p \left( N^{-\frac12} \right).\]
Consider
\begin{align*} 
\log ( r_{0k}^{-n_k^{(0)}}  r_{1k}^{-n_k^{(1)}})  &= N[p_k \log(p_k)- \lambda_0 p_{0k} \log(p_{0k})- \lambda_1 p_{1k}\log(p_{1k})] \\
&= N\{ \pi_k^* \log(p_k)- \lambda_0 \pi_{k}^{(0)} \log(p_{0k})- \lambda_1 \pi_{k}^{(1)}\log(p_{1k})\} +O_p \left(N^{1/2}\right) \\ 
&= N\{ \pi_k^* \log(\pi_k^*)- \lambda_0 \pi_{k}^{(0)} \log(\pi_k^{(0)})- \lambda_1 \pi_{k}^{(1)}\log(\pi_{k}^{(1)})\} +O_p \left(N^{1/2}\right) \\
&= -N\Big\{ \lambda_0 \pi_{k}^{(0)} \log\Big(\frac{\pi_k^{(0)}}{\pi_k^*}\Big)+\lambda_1 \pi_{k}^{(1)} \log\Big(\frac{\pi_k^{(1)}}{\pi_k^*}\Big)\Big\} +O_p \left(N^{1/2}\right) .
\end{align*}
Thus, by Theorem 1, 
\[\frac{\pr (H_0|C^{(0)},C^{(1)})}{\pr (H_1|C^{(0)},C^{(1)})} \sim c  N^{\frac{K-1}{2}} \prod_{k=1}^K \left(\frac{\pi_k^{(0)}}{\pi_k^*}\right)^{-N\lambda_0\pi_k^{(0)}} \left(\frac{\pi_k^{(1)}}{\pi_k^*}\right)^{-N\lambda_1\pi_k^{(1)}} \exp \left\{O_p \left(N^{1/2}\right) \right\}.\]
\end{proof}

\subsection{Proof of Theorem~\ref{SingMixtureCont}}
\label{SingMixtureContProof}
\begin{proof}
The result follows from Corollary 2.1 of \cite{Kleijn2006}.  The space $\mathbb{F}$ is compact relative to total variation distance, and hence is bounded with respect to $d$.  Hence the covering numbers $N(\epsilon,\mathbb{F},d)$ are finite for all $\epsilon > 0$.  The space $\mathbb{F}$ is also convex, and so it follows from Lemmas 2.2 and 2.3 of \cite{Kleijn2006} that the entropy condition of Corollary 2.1 is satisfied for the metric $d$.   
\end{proof}

\subsection{Proof of Corollary~\ref{PiCons}}
\label{PiConsProof}
\begin{proof}
Fix $\epsilon>0$.  
Because  KL($f^*||f)$ is defined, $f^*$ and $f_0$ have common support.  Therefore,
$d(f,f^*) = 0$ implies H$(f,f^*)=0$, where $H$ is the Hellinger distance
\[H^2(f,f^*)=\frac{1}{2} \int (f^{1/2}-f^{* 1/2})^2.\]
Hence, d$(\sum \pi_k f_k,f^*)=0$ implies $f=f^*$, and therefore $\Pi=\Pi^*$ by the uniqueness assumption.  Because $d(\sum \pi_k f_k, f^*)$ is continuous with respect to $\Pi$, there exists $\delta>0$ such that   $d(\sum \pi_k f_k,f^*) \leq \delta$ implies $||\Pi-\Pi^*|| \leq \epsilon$.  Therefore, 
\[\pr (\Pi \in \mathbb{S}^{K-1}: ||\Pi-\Pi^*|| < \epsilon \mid X) \leq \pr\{ f \in \mathbb{F}: d(f,f^*) > \delta \mid X\} \rightarrow 0\]
by Theorem~\ref{SingMixtureCont}. 
\end{proof}

\subsection{Proof of Remark~\ref{fstarcond}}
\label{fstarcondproof}
\begin{proof}
As $\KL (f_0||\sum \pi_kf_k)$ is globally convex with respect to $\Pi$, the minimum divergence is achieved when all first-order derivatives are $0$. Fix $\pi_3,\hdots,\pi_K$ and let $\pi_1 = a$, $\pi_2 = 1-a-\sum_{k=3}^K \pi_k$ for $0 \leq a \leq 1-\sum_{k=3}^K \pi_k$.  Let 
\[f^{(a)} = af_1+\Big( 1-a-\sum_{k=3}^K \pi_k\Big) f_2+\sum_{k=3}^K \pi_kf_k.\]
Then
\begin{align*}
\frac{\partial}{\partial a} \KL (f_0||f^{(a)}) &=-\int \frac{\partial}{\partial a} \log(f^{(a)})f_0 = -\int \frac{f_1}{f^{(a)}} f_0+\int \frac{f_2}{f^{(a)}} f_0.
\end{align*}
Hence, $\partial \KL (f_0||f^{(a)}) / \partial a=0$ implies that \[\int \frac{f_1}{f^{(a)}} f_0=\int \frac{f_2}{f^{(a)}} f_0.\]
An analogous result holds for any $\pi_i,\pi_j$ with $i\neq j$.  Therefore, if $f^*=\underset{f \in \mathbb{F}}{\text{argmin}} \, \text{KL}(f_0|| f)$ with $\pi_k^* > 0$ for all $k$, 
\[\int \frac{f_1}{f^*} f_0 = \cdots=  \int \frac{f_K}{f^*} f_0.\]

If $\pi_k^* = 0$ for some $k$, a similar argument shows that $\int (f_k/f^*) f_0$ must be equivalent for all $\pi_k^*>0$. 
\end{proof}

\subsection{Proof of Theorem~\ref{ConsTheorem}}
\label{ConsTheoremProof}
\begin{proof}
Let $C$ indicate group membership, so that the generative distribution for $x_n \in \{X^{(0)},X^{(1)}\}$ is 
\[g(f^{(0)},f^{(1)},C) \sim  \left\{ \begin{array}{c} f^{(0)},  C = 0, \\ f^{(1)},  C = 1.  \end{array} \right.\]
Note that
\begin{align*} 
\KL (g(\hat{f}^{(0)},\hat{f}^{(1)},C)||g(f^{(0)},f^{(1)},C)) &= \int (1-C) f^{(0)} \log{\frac{f^{(0)}}{\hat{f}^{(0)}}}+\int C f^{(1)} \log{\frac{f^{(1)}}{\hat{f}^{(1)}}} \\
&=\lambda_0 \KL (\hat{f}^{(0)}||f^{(0)})+(1-\lambda_0) \KL (\hat{f}^{(1)}||f^{(1)}).
\end{align*}
So, for $(\hat{f}^{(0)},\hat{f}^{(1)}) \in \mathbb{F}^2$ the divergence with the generative distribution is minimized at  $\hat{f}^{(0)}=f^{* (0)}$ and $\hat{f}^{(1)}=f^{* (1)}$.  As $P_0<1$, the prior has positive support over $\mathbb{F}^2$ and therefore the concentration conditions of Theorem~\ref{SingMixtureCont} are satisfied.  It follows from Corollary~\ref{PiCons} that \begin{align} \pr (||\hat{\Pi}^{(0)}-\Pi^{* (0)}|| \geq \epsilon \mid X) \rightarrow 0 \, \, , \, \,   \pr (||\hat{\Pi}^{(1)}-\Pi^{* (1)}|| \geq \epsilon \mid X) \rightarrow 0, \, \, \epsilon>0.\label{PiConvEq}\end{align} 

Assume $f^{* (0)} \neq f^{* (1)}$ and fix $\epsilon < ||\Pi^{* (0)}-\Pi^{* (1)}||$. From (\ref{PiConvEq}),  
$\pr (||\hat{\Pi}^{(0)}-\hat{\Pi}^{* (1)}|| < \epsilon \mid X) \rightarrow 0.$
This implies that $\pr (H_0 \mid X) \rightarrow 0$, as $\pr (H_0 \mid X) < \pr (||\hat{\Pi}^{(0)}-\hat{\Pi}^{* (1)}||< \epsilon \mid X)$. 

Assume $f^{* (0)} = f^{* (1)}=f^*$, where $f^*$ has weights $\Pi^*$.  Let 
\[A_\delta = \{\Pi^{(0)},\Pi^{(1)}: \, \, ||\Pi^{(0)}-\Pi^*|| < \delta \, \, , \, \, ||\Pi^{(1)}-\Pi^*|| < \delta \}. \]
Let $f_\alpha$ be the density for a Dir($\alpha$) distribution and $f(x|\Pi) = \sum_{k=1}^K \pi_k f_k$.  For large $N$,
\begin{align*}
\pr (A_{\delta}, X \mid H_1) &= \underset{\Pi^{(0)},\Pi^{(1)} \in A_{\delta}}{\iint} \prod_{i=1}^{N_0} f(x_i \mid \Pi^{(0)})  \prod_{j=1}^{N_1} f(x_j \mid \Pi^{(1)})  f_\alpha (\Pi^{(0)}) f_\alpha (\Pi^{(1)}) \\
& \leq  \underset{\Pi^{(0)},\Pi^{(1)} \in A_{\delta}}{\iint} \prod_{i=1}^{N_0} f(x_i \mid \Pi^{(0)})  \prod_{j=1}^{N_1} f(x_j\mid\Pi^{(0)})  f_\alpha (\Pi^{(0)}) f_\alpha (\Pi^{(1)}) \\
&=\pr (A_{\delta}, X \mid H_0)  \pr (A_{\delta} \mid H_0), 
\end{align*}
and so 
\begin{align*} 
\pr (H_1 \mid A_\delta,X) &= \frac{\pr (H_1) \pr (A_{\delta}, X \mid H_1)}{\pr (H_1) \pr (A_{\delta}, X \mid  H_1)+P_0\pr (A_{\delta}, X \mid H_0) } \\
&\leq \frac{\pr (H_1)\pr (A_\delta \mid H_0)}{\pr (H_1)\pr (A_\delta \mid H_0)+P_0}.
\end{align*}
Clearly $\pr (A_\delta \mid H_0) \rightarrow 0$ as $\delta \rightarrow 0$, and therefore 
\begin{align}
\pr (H_1 \mid A_\delta,X) \rightarrow 0 \,, \, \, \, \delta \rightarrow 0. \label{H0conv}
\end{align}  Result (\ref{PiConvEq}) implies that for all $\delta>0$, 
\begin{align}
\pr (\bar{A}_\delta \mid X) \rightarrow 0 \,, \, \, \, N \rightarrow 0. \label{AepConv}
\end{align} 

Fix $\epsilon>0$.  It follows from (\ref{H0conv}) and (\ref{AepConv}) that we may take $\delta$ sufficiently small to ensure that \[\pr (H_1 \mid X) = \pr (\bar{A}_\delta \mid X) \pr (H_1 \mid \bar{A}_{\delta} X)+  \pr (A_\delta \mid X) \pr (H_1 \mid A_{\delta} X) < \epsilon \]
for large $N$.  Therefore, $\pr (H_0 \mid X) \rightarrow 1$ as $N \rightarrow \infty$.   
\end{proof}

\section{Pseudocode}
\label{Pseudocode}
Here we give the details of the estimation procedure for the application to TCGA data, as introduced in Section 5 of the main article. To estimate the kernel parameters $\theta_k = (\mu_k, 1/\sigma_k^2)$ we use the flexible normal-gamma prior  
\[ \text{NG}(\mu_0=0.5, \lambda_0=1,a_0=1,b_0=0.5).\]
After randomly selecting a subsample of 500 CPG sites, the kernels are estimated via Gibbs sampling, where each iteration proceeds as follows:
\begin{itemize}
\item Draw $C^{(0)},C^{(1)} \mid  \Pi,X, \Theta$. The probability that for variable $m$ subject $n$ is allocated to kernel $k$ is 
 \[\pr (C_{mn}=k\mid X^{(i)}, \Pi^{(i)}, \Theta) \propto \pi_k f(X_{mn} \mid \mu_k,\sigma_k,[0,1]),\]
where $f(\cdot \mid \mu_k,\sigma_k,[0,1])$ is the density of a normal distribution with mean $\mu_k$ and standard deviation $\sigma_k$, truncated to fall within the interval $[0,1]$.  

\item Draw $\{\Pi_m\}_{m=1}^M\mid  \mathbb{C}^{(0)},\mathbb{C}^{(1)}$, where $\tilde{\pi}_{mk} \sim \text{Dir}(\alpha+\vec{n}_m)$.
\[\Pi^{(i)} = \pr (H_0\mid C^{(0)},C^{(1)}) \tilde{\Pi}_k+(1-\pr (H_0\mid C^{(0)},C^{(1)})) \tilde{\Pi}_k^{(i)}.\]
\item Draw $\Theta \mid C,X$.  The posterior distribution for $\theta_k$, $k=1,\hdots,K$ is
\[\theta_{k} \sim \text{NG}(\mu_{0k}, \lambda_{0k}=1,a_{0k},b_{0k}).\]
Let $X_k$ be the collection of values, over all probes, belonging to kernel $k$.   To account for truncation, generate $\mathbbm{Y}_k = \tilde{F}_k^{-1}F_k(X_k)$, where $\tilde{F}_k$ is the CDF for N$(\mu_k,\sigma_k)$ without truncation and $F_k$ is the CDF with truncation. Let $N_{k}$ be the total number of values belonging to kernel $k$, $\bar{Y}_{k}$ be the sample mean of $\mathbbm{Y}_k$, and $S_{k}$ the sample variance for $\mathbbm{Y}_k$. The posterior normal-gamma parameters are 
\begin{itemize}
\item $\mu_{0k} = \frac{\lambda_0 \mu_{0} + N_{k} \bar{Y}_{k}}{\lambda_0+N_{k}}$ 
\item  $\lambda_{0k} = \lambda_0+N_{k}$ 
\item $a_{0k} =a_{0}+\frac{N_k}{2}$
\item $b_{0k} = b_{0}+\frac{N_{k}S_{k}}{2}+\frac{\lambda_0 N_{k}(\bar{Y}_{k}-\mu_{0})^2}{2(\lambda_0+N_{k})}$.
\end{itemize}
\item Estimate $\alpha$ from $\{\Pi_m\}_{m=1}^M$ as in \cite{Ronning1989}.
\end{itemize}
For each Gibbs iteration we relabel kernels if necessary to maintain the order $\mu_1 < \mu_2 < \hdots < \mu_K$.  We average over the resulting Gibbs samples to obtain point estimates for $\{\theta_k\}_{k=1}^K$ and $\alpha$.  

For two-class testing, we put a uniform Be$(a_0=1,b_0=1)$ prior on $P_0$, and Gibbs sample as follows: 
\begin{itemize}
\item Draw $C^{(0)},C^{(1)} |  \{\Pi_m^{(0)},\Pi_m^{(1)}\}_{m=1}^M,X^{(0)},X^{(1)}$   The probability that for variable $m$ subject $n$ in class $i$ is realized from component $k$ is 
 \[\pr (C_{mn}^{(i)}=k\mid X_m^{(i)}, \Pi_m^{(i)}) \propto \pi_{mk}^{(i)} f(X_{mn}^{(i)} | \mu_k,\sigma_k,[0,1]).\] 
\item Compute $p_m=\pr (H_{0m} \mid \mathbb{C}_m^{(0)},\mathbb{C}_m^{(1)})$ as in Section 2.2,  Equation (2) for $m=1,\hdots,M$.
\item Draw $\{\Pi_m^{(0)},\Pi_m^{(1)}\}_{m=1}^M\|\mathbb{C}^{(0)}, \mathbb{C}^{(1)},P_0$.   For $\tilde{\Pi}_k \sim \text{Dir}(\alpha+\vec{n})$,  $\tilde{\Pi}_k^{(0)} \sim \text{Dir}(\alpha+\vec{n}^{(0)})$, $\tilde{\Pi}_k^{(1)} \sim \text{Dir}(\alpha+\vec{n}^{(1)})$, for class $i=0,1$
\[\Pi_m^{(i)} = P_m \tilde{\Pi}_k+(1-P_m) \tilde{\Pi}_k^{(i)}.\]

\item Draw $P_0$ from Be$(1+\sum_{m=1}^M P_m,1+M-\sum_{m=1}^M P_m)$. 
\end{itemize}

\section{Likelihood cross-validation}
\label{LikeCrossVal}

For the application to methylation array data described in Section 5 of the main article, we choose the number of dictionary kernels based on the mean log-likelihood for held out observations.  Here we describe this process in more detail and illustrate the results.  We also compare with results for kernels that are estimated independently at each CpG site, rather than shared across sites

\begin{figure}[!ht]
	\centerline{\includegraphics[scale=0.8, trim = 0mm 0mm 0mm 0mm, clip = TRUE]{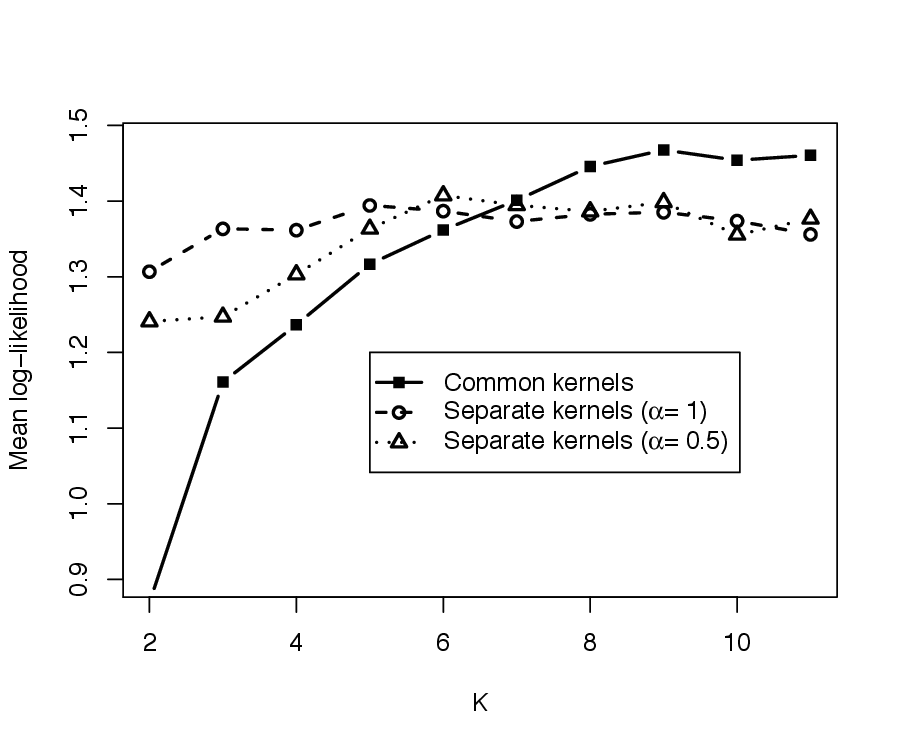}} 
	\caption{The mean cross-validated log-likelihood for held-out samples across CpG sites is shown when a common set of kernels is estimated across CpG sites, and when kernels are estimated separately for each site.}
	\label{crossval}
\end{figure}

 For each $K$, we estimate a dictionary of $K$ kernels from a sub-sample of 500 CpG sites, as described in Section~\ref{Pseudocode}.  Then, we select a separate sub-sample of 500 sites for cross-validation.  For each site, we randomly select a sample to hold out and estimate the kernel weights for that site based on the $N-1$ remaining samples; we then compute the log-density for the held out sample using the estimated kernel weights.  The mean log-density among the 500 held-out samples is shown in Figure~\ref{crossval} for $K=2,\hdots,11$.  The cross-validated likelihood is maximized at $K=9$.
 
For comparison, Figure~\ref{crossval} also displays the cross-validated likelihood when the kernels are estimated independently for each site, using standard choices for the Dirichlet concentration parameter $\alpha = 0.5$ and $\alpha=1$.  It is not surprising that independent estimation is superior when $K$ is too small for the shared kernel model to adequately capture the diversity of distributions across CpG sites.  But for adequately large $K$ the shared kernel model gives superior results.  This is evidence that the use of common kernels across sites is not only computationally convenient but is also advantages due to the borrowing of information.

\section{Simulation studies}

\subsection{Bayes factor convergence}
\label{SimSetup}

We investigate the performance of the shared kernel test for simulated data with large $N$, and assess the asymptotic rates derived in Section 3 of the main article under more general conditions.  We simulate $200$ separate univariate datasets from the assumed model as follows:
\begin{enumerate}
\item Draw $N$ uniformly on a log-scale from $10$ to $1,000,000$.
\item Draw $K$ uniformly from $\{2,\hdots,9\}$.
\item Draw $\mu_1,\hdots,\mu_K$ independently from Un$(0,1)$.
\item Draw $\sigma_1,\hdots,\sigma_K$ independently from Un$(0,\frac{1}{K})$
\item Draw $H_0$ from Bernoulli$(0.5)$
\item If $H_0=1$
\begin{itemize}
\item Draw $\Pi$ from a uniform, $K$-dimensional Dirichlet distribution
\item For $n=1,\hdots,N$ assign $x_n$ to either class 0 or class 1 with equal probability.  
\item Draw $x_1,\hdots,x_N \in \mathbb{X}$ from density
\[\sum_{k=1}^K \pi_k \text{Tnorm}(\mu_k,\sigma_k,[0,1]),\]
where Tnorm defines the density of a truncated normal distribution.
\end{itemize}
\item If $H_0=0$
\begin{itemize}
\item  Draw $\Pi^{(0)}$ and $\Pi^{(1)}$ independently from a uniform, $K$-dimensional Dirichlet distribution
\item For $n=1,\hdots,N$ assign $x_n$ to either class $0$ or class $1$ with equal probability.  
\item Draw $x_1,\hdots,x_{N_0} \in \mathbb{X}^{(0)}$ from
\[\sum_{k=1}^K \pi_k^{(0)} \text{Tnorm}(\mu_k,\sigma_k,[0,1])\]
\item Draw $x_1,\hdots,x_{N_1} \in \mathbb{X}^{(1)}$ from
\[\sum_{k=1}^K \pi_k^{(1)} \text{Tnorm}(\mu_k,\sigma_k,[0,1]).\]
\end{itemize}
\end{enumerate}

For each simulation, we estimate the posterior for both the component distributions and group weights simultaneously.  We fix $P_0=0.5$ and use a truncated normal-gamma prior for the component densities, as described in Section~\ref{Pseudocode}.  

\begin{figure}[!ht]
	\centerline{\includegraphics[scale=0.5, trim = 0mm 0mm 0mm 0mm, clip = TRUE]{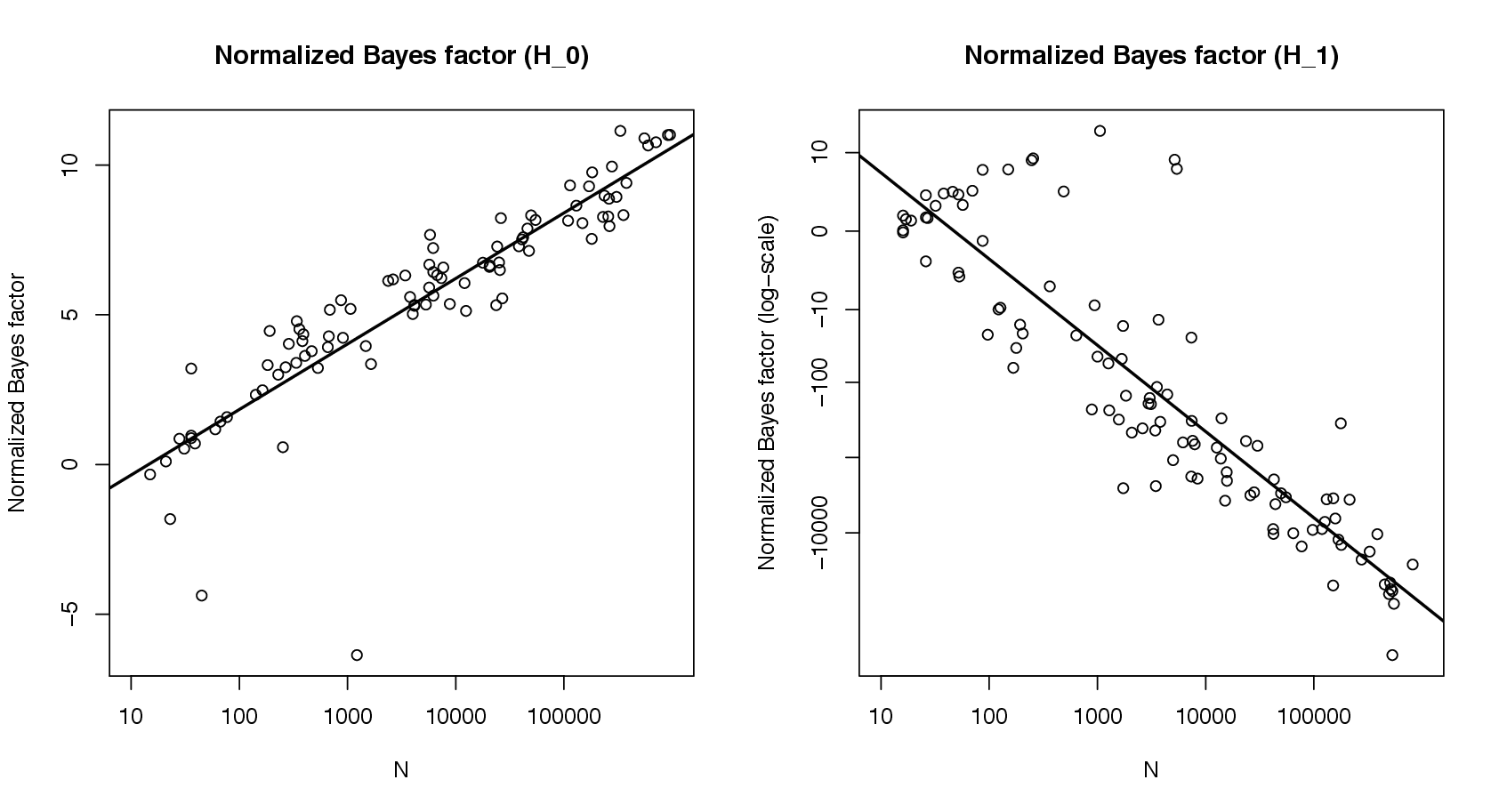}} 
	\caption{Normalized Bayes factors (see (\ref{h0norm}) and (\ref{h1norm})) versus $N$, under $H_0$ (left plot) and $H_1$ (right plot). Under $H_0$ the normalized Bayes factors are given on the original scale, under $H_1$ they are given on a log scale.  Both plots show a linear trend.  } 	\label{NormBayesFacts}
\end{figure}

To investigate the behavior of the posterior probability of $H_0$ as a function of $N$, we normalize the log of the posterior Bayes factor as suggested by the dominating term from the asymptotic rates in Section 3 of the main article.  Specifically, under $H_0$ the normalized Bayes factor is 
\begin{align} \frac{2}{K-1} \log \left \{ \frac{\pr (H_0|X)}{\pr (H_1|X)} \right\} \label{h0norm} \end{align}
and under $H_1$ the normalized Bayes factor is 
\begin{align} \frac{\log \left \{ \frac{\pr (H_0|X)}{\pr (H_1|X)} \right\} }{\sum \Big\{ \lambda_0 \pi_{k}^{(0)} \log\Big( \frac{\pi_k^{(0)}}{\pi_k^*}\Big)+(1-\lambda_0) \pi_{k}^{(1)} \log\Big( \frac{\pi_k^{(1)}}{\pi_k^*}\Big) \Big\} }. \label{h1norm}\end{align}
Figure~\ref{NormBayesFacts} shows the normalized Bayes factor for each of $200$ simulations.  As expected, $\pr (H_0|X)$ tends to $1$ under $H_0$ and tends to $0$ under $H_1$, as $N \rightarrow \infty$.  Moreover, the log of the Bayes factor tends to $-\infty$ at an approximately linear rate with $N$ under $H_1$, and tends to $+\infty$ at an approximately log-linear rate with $N$ under $H_0$.  These rates agree with the asymptotic forms derived in Section 3 of the main article.  

\subsection{Distribution recovery}
\label{DistRec}
Here we investigate the ability of the two-class method to recover the generative distribution.  We compare distribution recovery under the two-class model versus 
\begin{enumerate}
	\item Fitting a Dirichlet mixture model separately for each class, and
	\item Fitting a Dirichlet mixture model that ignores class distinctions.  
\end{enumerate}
Ideally, we would like the two-class model to have similar performance to approach 1 under $H_1$ and similar performance to approach 2 under $H_0$.  

\begin{figure}[!ht]
	\centerline{\includegraphics[scale=0.5, trim = 0mm 0mm 0mm 0mm, clip = TRUE]{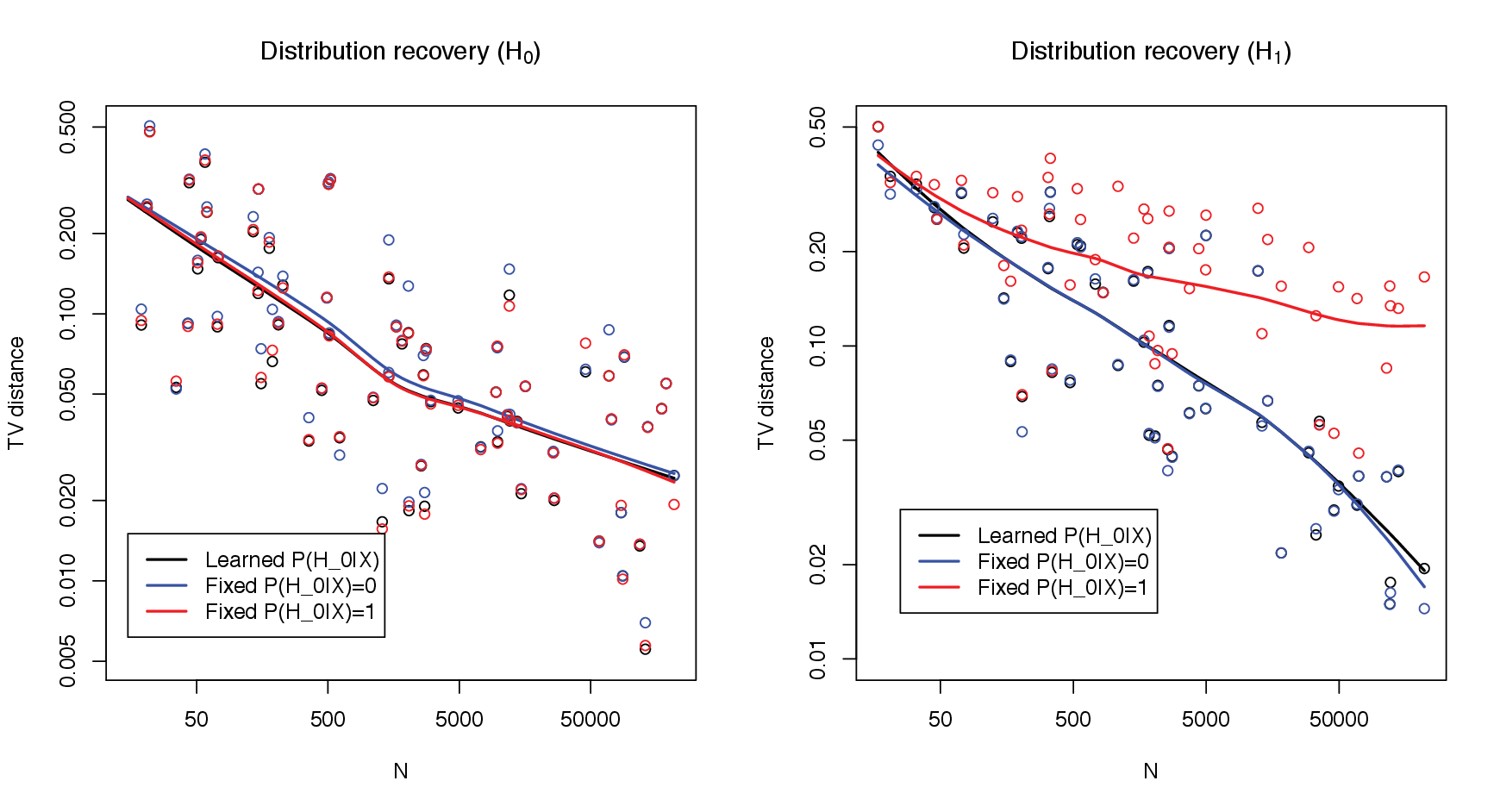}} 
	\caption{Total variation distance between the estimated posterior and the generative model for $200$ randomly generated simulations under $H_0$ (left plot) and $H_1$ (right plot).  Results are shown for the two-class model (learned $P(H_0|X)$), separate estimation of each class (fixed $P(H_0|X)=0$), and common estimation ignoring class labels (fixed $P(H_0|X)=1$).}
	\label{distrecov}
\end{figure}

We simulate $200$ univariate datasets and estimate the posterior for the two-class model as in Section~\ref{SimSetup}.  Separate and common models (approaches 1 and 2 above) are estimated analogously except for the dependence between classes.  Figure~\ref{distrecov} shows the total variation distance between the mean posterior distribution and the generative distribution for each simulation and using each of the three estimation approaches. Not surprisingly, separate estimation performs much better under $H_1$ and common estimation performs marginally better under $H_0$.  The flexible two-class approach performs similarly to common estimation under $H_0$ and separate estimation under $H_1$, even for smaller sample sets.

\subsection{Posterior probability performance}
\label{PostProbPerf}

Here we describe an extensive simulation that concerns the accuracy and precision of estimates for the posterior probability of equality between groups.  This simulation illustrates the advantage of estimating shared kernels and shared prior probability across variables, relative to the number of variables $M$ and sample size $N$.  In this example data are not truncated between $0$ and $1$, and so kernels are estimated using a standard normal-gamma mixture. 

For a given number of variables $M$, sample size $N$, and proportion of variables with no group difference $P$, data were simulated from a mixture of $5$ Gaussian kernels as follows:
\begin{itemize}
\item Draw $\mu_1,\hdots,\mu_5$ independently from $\text{Ga}(1,1)$.
\item Draw $\sigma_1,\hdots,\sigma_5$ independently from $\text{Un}(0,1/2)$.
\item For variables $m=1$ through $m=PM$, draw data under $H_0$
\begin{itemize}
\item Draw $\Pi$ from a uniform Dirichlet distribution
\item Draw $x_{m1},\hdots,x_{mN}$ from $\sum_{k=1}^K \pi_k \text{N}(\mu_k,\sigma_k).$
\end{itemize}
\item For variables $m=P M+1$ through $m=M$, draw data for two groups of size $N/2$
\begin{itemize}
\item Draw $\Pi^{(0)}$ and $\Pi^{(1)}$ independently from a uniform Dirichlet distribution
\item Draw $x_{m1},\hdots,x_{m (N/2)}$ from $\sum_{k=1}^K \pi_k \text{N}(\mu_k,\sigma_k).$
\item Draw $x_{m (N/2+1)},\hdots,x_{mN}$ from $\sum_{k=1}^K \pi_k \text{N}(\mu_k,\sigma_k).$
\end{itemize}
\end{itemize}
Generation of kernel means and standard deviations differ substantially from the prior assumption of normal means and inverse-gamma variances. Five repeated simulations were performed for each combination of the values $M=\{10,60,360\}$, $N=\{30, 120,480\}$, and $P = \{0.1,0.2,\hdots,0.9\}$.  For each simulation we estimated the posterior probability of $H_0$ for each variable and computed the Bayes error
\[\sum_{m=1}^M [\{1-\mathbbm{1}(H_{0m})\} \pr (H_{0m} \mid X ) +  \mathbbm{1}(H_{0m}) \{1- \pr (H_{0m} \mid X )\}]/M.\]

Posterior probabilities were estimated using four approaches: with shared kernels and shared estimate for $P_0$ among variables, with shared kernels among variables and fixed $P_0=0.5$, with independently estimated kernels and fixed $P_0=0.5$, and using the co-OPT method \citep{Ma2011} under default specifications.  Table~\ref{tab3} shows the mean Bayes error over all simulations for the given values of $M$ and $N$.  The co-OPT test is included for reference and generally does not perform as well as the other three methods.  The incorporation of shared kernels among variables and the incorporation of a shared estimate for $P_0$ also both generally reduced error.  The relative benefit of borrowing information across variables in the form of shared kernels or shared $P_0$ was (unsurprisingly) more dramatic for large $M$ and small $N$.

\begin{table}[!ht]
\renewcommand{\arraystretch}{0.8}
\begin{tabular}{c| l c c c}
 & & $\mathbf{M=10}$ & $\mathbf{M=60}$ & $\mathbf{M=360}$ \\
\hline
 \multirow{4}{*}{$\mathbf{N=30}$} & Shared kernels and estimated $P_0$  & $\mathbf{0.40} \pm 0.03$ & $\mathbf{0.32} \pm 0.02$ & $\mathbf{0.31} \pm 0.02$\\
 &  Shared kernels and $P_0=0.5$  & $\mathbf{0.41} \pm 0.02$ & $\mathbf{0.36} \pm 0.02$ & $\mathbf{0.36} \pm 0.01$ \\
 &  Separate kernels and $P_0=0.5$  & $\mathbf{0.47} \pm 0.02$ & $\mathbf{0.47} \pm 0.01$ & $\mathbf{0.47} \pm 0.01$\\
 &  co-OPT test & $\mathbf{0.46} \pm 0.02$ & $\mathbf{0.49} \pm 0.01$ & $\mathbf{0.49} \pm 0.02$\\
\hline
 \multirow{4}{*}{$\mathbf{N=120}$} & Shared kernels and estimated $P_0$  & $\mathbf{0.20} \pm 0.04$ & $\mathbf{0.19} \pm 0.03$ & $\mathbf{0.16} \pm 0.01$\\
 &  Shared kernels and $P_0=0.5$  & $\mathbf{0.20} \pm 0.03$ & $\mathbf{0.20} \pm 0.02$ & $\mathbf{0.18} \pm 0.01$ \\
 &  Separate kernels and $P_0=0.5$  & $\mathbf{0.32} \pm 0.02$ & $\mathbf{0.30} \pm 0.04$ & $\mathbf{0.30} \pm 0.01$\\
 &  co-OPT test & $\mathbf{0.40} \pm 0.02$ & $\mathbf{0.40} \pm 0.02$ & $\mathbf{0.43} \pm 0.03$\\
\hline
 \multirow{4}{*}{$\mathbf{N=480}$} & Shared kernels and estimated $P_0$  & $\mathbf{0.07} \pm 0.02$ & $\mathbf{0.09} \pm 0.02$ & $\mathbf{0.08} \pm 0.01$\\
 &  Shared kernels and $P_0=0.5$  & $\mathbf{0.08} \pm 0.02$ & $\mathbf{0.09} \pm 0.02$ & $\mathbf{0.09} \pm 0.01$ \\
 &  Separate kernels and $P_0=0.5$  & $\mathbf{0.12} \pm 0.05$ & $\mathbf{0.14} \pm 0.02$ & $\mathbf{0.13} \pm 0.01$\\
 &  co-OPT test & $\mathbf{0.29} \pm 0.07$ & $\mathbf{0.28} \pm 0.03$ & $\mathbf{0.29} \pm 0.04$\\
\hline 
\end{tabular}
\\
\caption{Mean Bayes errors under four testing approaches.  For a given number of variables $M$ and sample size $N$,  the mean is computed from $45$ simulations: $5$ repeated simulations for $P=0.1,\hdots,0.9$, where $P$ is the proportion of variables with no group difference.  The given margin of error in each cell is twice the standard error for the mean.}
\label{tab3}
\end{table}

\begin{figure}[!ht]
	\centerline{\includegraphics[scale=1, trim = 0mm 0mm 0mm 0mm, clip = TRUE]{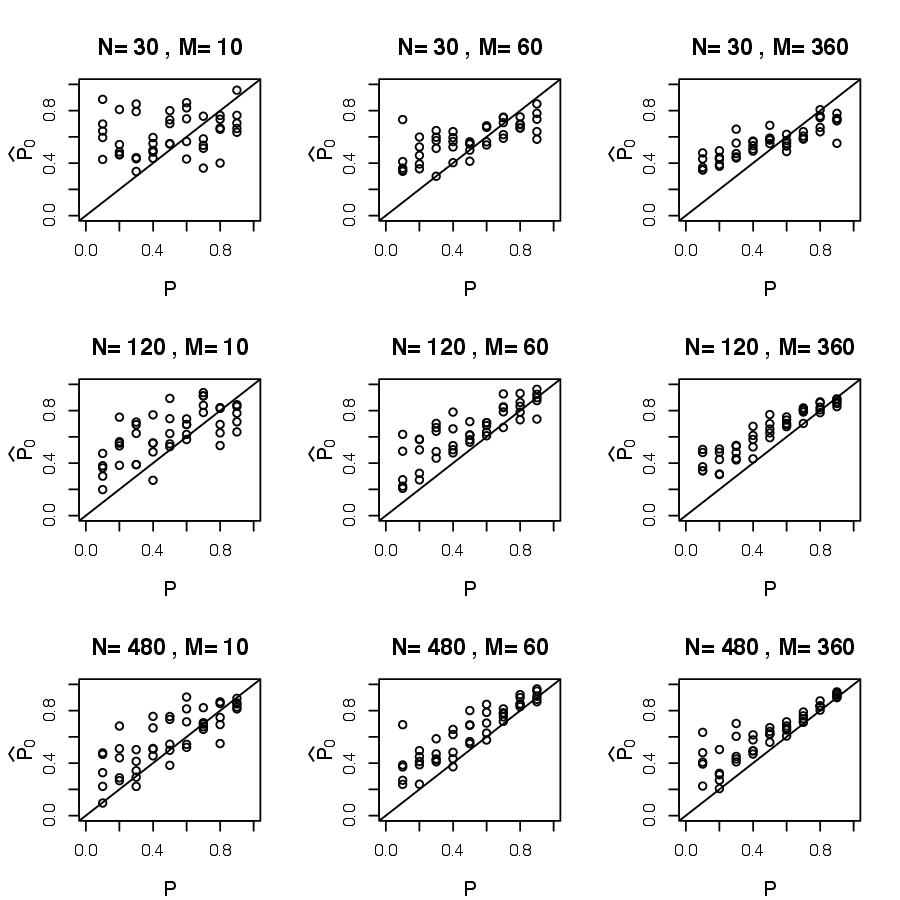}} 
	\caption{Estimated prior probabilities of no difference ($\hat{P}_0$) for each simulation versus the actual proportion of variables with no difference ($P$), for a given number of variables $M$ and sample size $N$.}
	\label{notruncpas}
\end{figure}

Figure~\ref{notruncpas} displays the estimated prior probabilities of no difference ($\hat{P}_0$) for each simulation versus the actual proportion of variables with no difference ($P$).  Estimates of $P_0$ are more accurate and precise for larger $M$ and larger $N$.  However, generally  $\hat{P}_0>P$, and this reflects the tendency of the prior to favor the null unless there is substantial evidence for the alternative.        

\bibliographystyle{biometrika}
\bibliography{MethylationTestingPaperBib}

\end{document}